\documentstyle[twocolumn,prb,aps,floats,epsfig]{revtex}

\begin{document}
\draft
\twocolumn[\hsize\textwidth\columnwidth\hsize\csname@twocolumnfalse\endcsname

\title{New magnetic phase in metallic V$_{2-y}$O$_3$ close to the metal
 insulator transition}

\author{S. Klimm, M. Herz, R. Horny, G. Obermeier, M. Klemm, and S. Horn}
\address{Universit\"{a}t Augsburg, 86135 Augsburg, Germany}
\date{\today}
\maketitle

\begin{abstract}

We have observed two spin density wave (SDW) phases in hole doped metallic
V$_{2-y}$O$_3$, one evolves from the other as a function of doping, pressure or
temperature. They differ in their response to an external magnetic field, which
can also induce a transition between them. The phase boundary between these two
states in the temperature-, doping-, and pressure-dependent phase diagram has
been determined by magnetization and magnetotransport measurements. One phase
exists at high doping level and has already been described in the literature.
The second phase is found in a small parameter range close to the boundary to
the antiferromagnetic insulating phase (AFI). The quantum phase transitions
between these states as a function of pressure and doping and the respective
metamagnetic behavior observed in these phases are discussed in the light of
structurally induced changes of the band structure.

\end{abstract}

\pacs{PACS numbers: 75.30.Fv, 75.30.Kz, 72.15.Gd, 71.30.+h}

\vskip3.0pc ]

\narrowtext

\section{Introduction}

\begin{figure}[bt]
 \begin{center}
  \epsfig{file=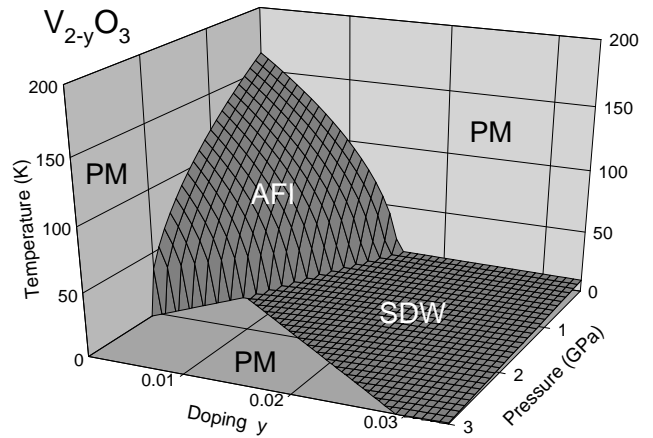, width=\columnwidth}
 \end{center}
 \caption{3-dimensional sketch of the phase diagram of V$_{2-y}$O$_3$ as a
  function of temperature, pressure, and doping.}
 \label{3D}
\end{figure}

The compound V$_{2-y}$O$_3$ shows a complicated phase diagram as a function of
temperature, pressure, oxygen stoichiometry and doping on the vanadium site
\cite{Imada98}. The pure compound exhibits a metal-insulator (MI) transition at
$T_{MI}=170$~K, which separates a metallic paramagnetic from an
antiferromagnetic insulating phase. The transition is shifted to lower
temperatures by external pressure, Ti doping on the vanadium site and vanadium
deficiency $y$, while Cr or Al doping increases $T_{MI}$. At sufficiently high
Cr and Al doping levels a paramagnetic insulating phase exists at room
temperature, which orders antiferromagnetically at $T_{AF}=200$~K. For vanadium
deficiencies $y \ge 0.015$, a spin density wave (SDW) phase forms at $T_{SDW}$.
The ordering temperature $T_{SDW}$ is a function of doping slightly decreasing
from $T_{SDW} \approx 11$~K close to the AFI~--~SDW boundary down to $T_{SDW}
\approx 8$~K at $y=0.04$ \cite{Carter91b,Ueda80}. Neutron scattering
measurements revealed the SDW to be of an incommensurate spiral type with wave
vector $q=1.7$~c* along the hexagonal $c$-axis and the moments lying within the
ab-planes \cite{Bao93}. In stoichiometric samples $T_{MI}$ is reduced by the
application of hydrostatic pressure with a critical pressure of $p_c=2$~GPa for
the complete suppression of the AFI-phase. However, in contrast to doping, the
pressure stabilized metallic phase remains paramagnetic down to lowest
temperatures. In doped samples external pressure reduces $T_{SDW}$. The
critical pressure increases on doping. A 3-dimensional sketch of the phase
diagram is shown in Fig.~\ref{3D}.

Recently it was found \cite{Klimm_APS00} that two different SDW phases can be
distinguished as a function of pressure and vanadium deficiency $y$. This
indicates that the electronic structure of V$_{2-y}$O$_3$ is still sensitive to
pressure and vanadium deficiency in the metallic phase, most likely due to a
change of weight or character of the bands at the Fermi energy $E_F$. A close
inspection of the phase diagram of V$_2$O$_3$ in the vicinity of the phase
boundaries may provide an understanding of the important parameters dominating
the physics of this classical compound and also the possible role of electronic
correlations.

We first demonstrate the existence of two different SDW phases and then discuss
the stability of the various phases of V$_{2-y}$O$_3$ based on LDA band
structure calculations. We present magnetic susceptibility and electronic
transport measurements as a function of vanadium deficiency $y$, applied
pressure $p$ and magnetic field $B$ in the vicinity of the phase boundary
between the antiferromagnetically ordered insulating and the metallic SDW phase
and the boundary between the two different SDW phases.

The electrical resistivity of V$_{2-y}$O$_3$ for different levels of doping and
as a function of pressure has been investigated earlier
\cite{Carter91b,Ueda80,McWhan73,Carter93}, but a systematic investigation
including the dependence of the magnetotransport on the crystallographic
orientation, which we will show below is essential for a complete picture of
the transport properties, has not been done.

\section{Experimental}

Single crystals of V$_{2-y}$O$_3$ were grown by the chemical vapor transport
technique with TeCl$_4$ as transport agent. The stoichiometry was controlled by
the composition of the starting materials. X-ray diffraction on powdered single
crystals showed the V$_{2-y}$O$_3$ samples to be single phase. Since no
non-destructive method with sufficient accuracy to determine the exact
stoichiometry of the grown crystals was available, the doping dependence of
$T_{MI}$ or $T_{SDW}$ as depicted from literature \cite{Ueda80,Rosenbaum95} was
used to estimate the values of $y$. In addition the $c$-axis parameter of the
crystals, which varies almost linearly with $y$ (Ref.~\onlinecite{Ueda80}), was
measured yielding consistent results for the values of $y$. Although there is a
relatively large error in the absolute values of $y$, the relative position of
the various samples on a scale of doping can assumed to be correct. The
orientation of the crystals was determined by Laue diffraction. Magnetic
measurements were performed using a Quantum Design SQUID magnetometer.
Rectangular bars with a typical size of $2\times0.3\times0.3$~mm$^3$ were cut
from the crystals to perform transport measurements along well defined
crystallographic directions. The resistivity was measured using a standard
4-probe AC-technique. Experiments under quasi hydrostatic pressure were
performed in a CuBe piston-cylinder cell with a manganin gauge and n-pentane --
isoamyl (1:1) as a pressure transmitting medium.

\section{Results}

\subsection{Electrical resistivity}

\begin{figure}[bt]
 \begin{center}
  \epsfig{file=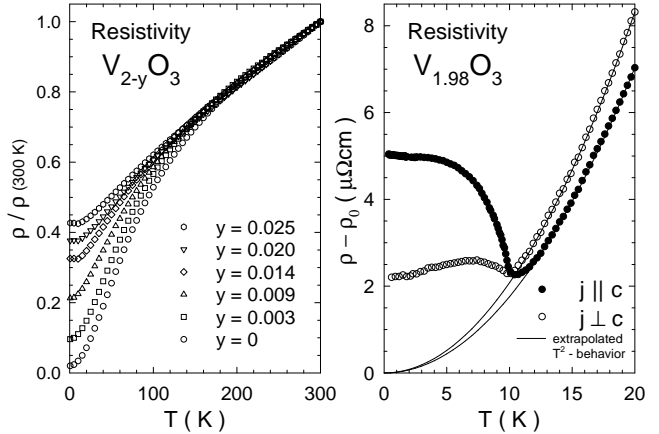, width=\columnwidth}
 \end{center}
 \caption{Left panel: Temperature dependence of resistivity for
  ${\bf j}\perp{\bf c}$ normalized to the value at $T=300$~K. The resistivity
  curves for $y<0.015$ are taken at a pressure just sufficient for the samples
  to remain metallic down to the lowest temperature.
  Right panel: Resistivity of V$_{1.98}$O$_3$ around $T_{SDW}$ with current
  along the two principal directions. The residual resistance $\rho_0$
  determined by a $T^2$-fit above $T_{SDW}$ is subtracted for better
  comparison.}
 \label{rho}
\end{figure}

The investigated samples of different stoichiometry will be labeled in the
following by the $y$ value estimated as described above. The temperature
dependent resistivity $\rho$ for different values of doping is shown in the
left part of Fig.~\ref{rho}. The curves for different $y$ are normalized to
their respective room temperature resistivity. The absolute value of the room
temperature resistivity ranges from 300 to 400~$\mu\Omega$cm. At ambient
pressure the stoichiometric sample ($y=0$) shows a MI-transition at
$T_{MI}=160$~K on cooling (175~K on warming up). Above the critical pressure at
$p=2$~GPa a $T^2$ dependence is observed at low temperatures in agreement with
existing literature \cite{Carter93,Carter94}. The residual resistance ratio of
$RRR=50$ shows the crystal to be of good quality. As expected the residual
resistivity $\rho_0$ increases monotonously with increasing vanadium vacancy
concentration. The shape of the resistivity curves at intermediate temperatures
depends on the current direction with respect to the crystallographic axes,
probably reflecting the anisotropic phonon spectrum in this compound
\cite{Yethiray87}. The negative curvature of resistivity is less pronounced for
${\bf j}\parallel{\bf c}$ (not shown here) than for ${\bf j}\perp{\bf c}$
resulting in a more linear temperature dependence \cite{Klimm97}.

For $y>0.015$ the transition into the SDW ground state at $T_{SDW}$ is marked
by a minimum in the resistivity for ${\bf j}\parallel{\bf c}$ (along the
propagation vector {\bf q} of the SDW), while for ${\bf j}\perp{\bf c}$ a less
pronounced kink is observed (Fig.~\ref{rho}, right panel) indicating the
formation of a gap along the $c^*$-direction in reciprocal space. The
temperature $T_{min}$ of the resistivity minimum for ${\bf j}\parallel{\bf c}$
as well as the kink for ${\bf j}\perp{\bf c}$ decreases only slightly on
application of an external magnetic field of up to $B=12$~T. The coefficient
$dT_{min}/dB \approx -0.1$~K/T is almost the same for ${\bf B}\parallel{\bf c}$
as well as for ${\bf B}\perp{\bf c}$ \cite{Klimm97}.

\subsection{Magnetic susceptibility}

\begin{figure}[bt]
 \begin{center}
  \epsfig{file=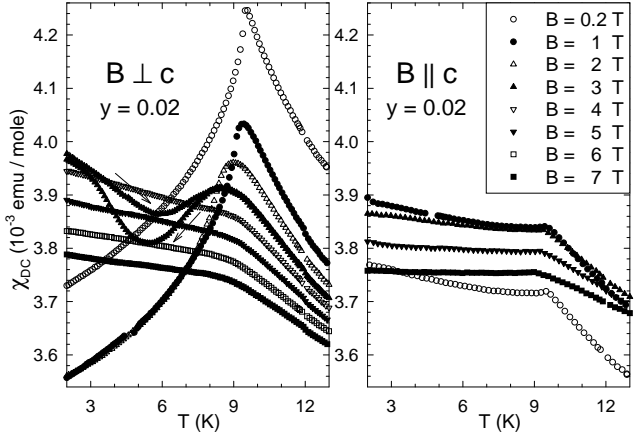, width=\columnwidth}
 \end{center}
 \caption{DC-susceptibility $M/B$ for a sample with $y=0.02$ at different
  magnetic fields applied perpendicular (left) and parallel (right) to the
  $c$-axis. The arrows at the ${\bf B}=3$~T curve indicate the warming
  and cooling cycle, respectively. Other curves do not show hysteresis.}
 \label{chi_1107}
\end{figure}

The response of the SDW state to an external magnetic field depends on the
orientation of the field with respect to the $c$-axis and on the level of
doping. For $y>0.018$ (deep in the SDW phase) the low field magnetic
susceptibility shows an antiferromagnetic cusp at $T_{SDW}$ for ${\bf
B}\perp{\bf c}$ while for ${\bf B}\parallel{\bf c}$ the susceptibility first
remains constant below $T_{SDW}$ and then even slightly increases towards lower
temperatures reflecting the response for the hard and easy AF axis,
respectively (Fig.~\ref{chi_1107}). This behavior is consistent with the in
plane orientation of the moments in this transverse polarized SDW phase and has
already been observed earlier \cite{Rosenbaum95}.

Only a weak field dependence of the susceptibility and a slight reduction of
$T_{SDW}$ with field is observed for ${\bf B}\parallel{\bf c}$. For ${\bf
B}\perp{\bf c}$, however, the cusp in the susceptibility as a function of
temperature, apparent at low fields, transforms into a kink above a critical
value $B_{c,\perp} \approx 3$~T. The slope at low temperatures changes from
positive to negative with increasing field indicating a reorientation of spins
above $B_{c,\perp}$. With an applied field of $B=3$~T, the transformation is
observed as a function of temperature with the susceptibility exhibiting a
considerable hysteresis, strongly suggesting a phase transition between two
magnetic phases.

\begin{figure}[bt]
 \begin{center}
  \epsfig{file=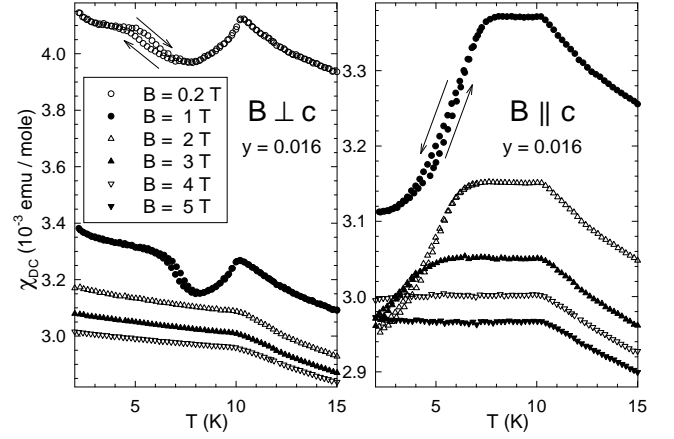, width=\columnwidth}
 \end{center}
 \caption{DC-susceptibility $M/B$ for a sample with composition close to the
  AFI phase ($y=0.016$) at different magnetic fields applied perpendicular
  (left) and parallel (right) to the $c$-axis.}
 \label{chi_316}
\end{figure}

Approaching the SDW~--~AFI phase boundary by reducing $y$, the critical field
$B_{c,\perp}$ decreases until for $y=0.016$ the magnetic transition is observed
even at the lowest fields (Fig.~\ref{chi_316}): Above 8~K the low field
susceptibilities for $y=0.016$ and $y=0.02$ (Fig.~\ref{chi_1107}) are similar.
Below 8~K, however, for $y=0.016$ the susceptibility for ${\bf B}\parallel{\bf
c}$ decreases abruptly while for ${\bf B}\perp{\bf c}$ it increases. Thus at
low temperatures and low doping the hard and easy axes appear interchanged when
compared to higher doping levels. For ${\bf B}\perp{\bf c}$ an increasing field
shifts the transition to higher temperatures, qualitatively similar to the
$y=0.02$ sample, but with the critical field shifted by $\approx 2.5$~T. A
magnetic field applied along the $c$-axis, however, shifts the transition of
the $y=0.016$ sample to lower temperatures (Fig.~\ref{chi_316}, right) and at
high fields a behavior as observed for the $y=0.02$ sample can be restored.

\begin{figure}[bt]
 \begin{center}
  \epsfig{file=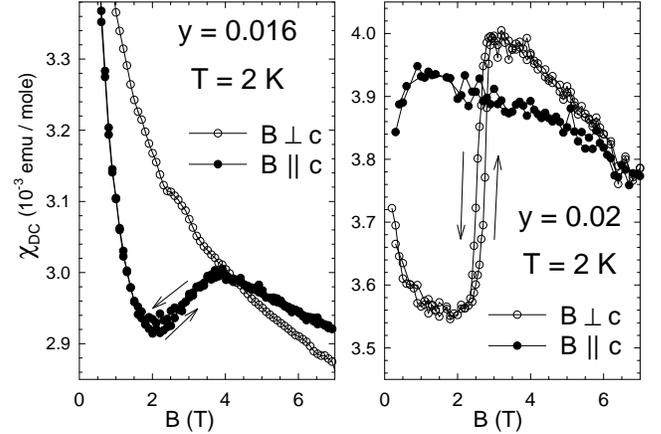, width=\columnwidth}
 \end{center}
 \caption{Field dependence of DC-susceptibility ($M/B$) in the SDW phase at
 $T=2$~K for the $y=0.016$ sample (left) and the $y=0.02$ sample (right).}
 \label{magnetization}
\end{figure}

Accordingly, the magnetization behaves differently as a function of applied
field in the two phases. This is most clearly seen plotting $\chi_{DC}=M/B$
vs.\ $B$ (Fig.~\ref{magnetization}). The right panel shows $\chi_{DC}$ of the
$y=0.02$ sample for {\bf B} parallel and perpendicular to the hexagonal
$c$-axis. For ${\bf B}\parallel{\bf c}$ $\chi_{DC}$ is almost independent of
$B$ while a metamagnetic transition at $B=B_{c,\perp}$ is observed for ${\bf
B}\perp{\bf c}$ in accordance with the $\chi(T)$ shown in Fig.~\ref{chi_1107}.
A similar behavior was reported by Bao {\it et al.} \cite{Bao01} for a sample
with an even higher vanadium deficiency ($y=0.04$) and was attributed to a spin
flop transition. The behavior of the $y=0.016$ sample is shown in the left
panel of Fig.~\ref{magnetization}. At $T=2$~K $\chi_{DC}(B)$ does not show a
transition for ${\bf B}\perp{\bf c}$ since it is already in the second phase.
But for ${\bf B}\parallel{\bf c}$ a transition occurs at a characteristic field
$B_{c,\parallel}$ as expected from the behavior of $\chi(T)$ shown in
Fig.~\ref{chi_316}.

The temperature at which a cusp or kink is observed in the susceptibility
tracks the temperature of the resistivity minimum as the applied magnetic field
is varied for both principal field directions. We therefore believe that the
metamagnetic behavior observed in the magnetization is not due to a suppression
of the SDW phase but rather to a reorientation of the moments or a change of
the propagation vector {\bf q}.

\begin{figure}[bt]
 \begin{center}
  \epsfig{file=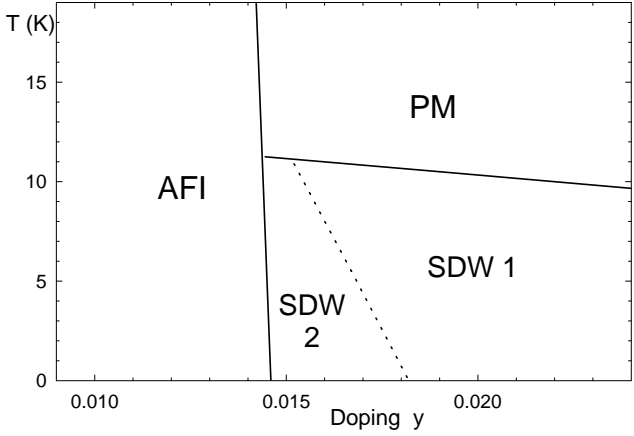, width=\columnwidth}
 \end{center}
 \caption{$T$~--~$y$ phase diagram for V$_{2-y}$O$_3$ including the new phase
  boundary (dotted line).}
 \label{T-y-pd}
\end{figure}

The results presented above suggest the existence of two different magnetic
phases within the SDW regime. In the following we will call these phases SDW~1
and SDW~2. SDW~1 designates the well known phase, already characterized by
neutron scattering \cite{Bao93}. The newly discovered phase, designated SDW~2,
exists in a small range of $y$ close to the SDW~--~AFI boundary, and is
inferred from anomalies observed in the magnetic susceptibility and
magnetoresistance. A transition from SDW~1 to SDW~2 with decreasing temperature
was observed for a crystal with $y=0.016$, implying that the phase boundary
between the two phases, in the temperature--doping diagram, has a negative
slope. Further evidence and a rough estimate for this negative slope is
provided by magnetoresistance and susceptibility measurements as a function of
magnetic field. A sample which is in the SDW~2 phase at $B=0$ transforms into
the SDW~1 phase, if a magnetic field above $B_{c,\parallel}$ is applied along
the $c$-axis. On the other hand, a field in excess of $B_{c,\perp}$ applied
perpendicular to the $c$-axis transforms the SDW~1 phase of a higher doped
sample into the SDW~2 phase. Both fields $B_{c,\perp}$ and $B_{c,\parallel}$
are monotonous functions of doping, i.e.\ they can be used as a measure of the
distance of a particular sample to the phase boundary between SDW~1 and SDW~2
phase (on the axis of doping). For given doping $B_{c,\parallel}$ decreases
while $B_{c,\perp}$ increases with increasing temperature in the vicinity of
this phase boundary. From this doping and temperature dependence of the
critical fields $B_{c,\perp}$ and $B_{c,\parallel}$ the slope of the phase
boundary between SDW~1 and SDW~2 in the $T$~--~$y$ plane can be roughly
determined as given in Fig.~\ref{T-y-pd}.

\subsection{Magnetoresistance}

The magnetoresistance (MR) shows a metamagnetic behavior corresponding to that
of the magnetic susceptibility. Since we can carry out MR measurements over a
larger range of temperature, magnetic field, and hydrostatic pressure we will
in the following concentrate on magnetoresistance rather than on magnetization
to further characterize the magnetic behavior of V$_{2-y}$O$_3$.

The low temperature region of the PM phase of stoichiometric samples ($p \ge
2$~GPa) exhibits a positive magnetoresistance with a quadratic field
dependence. The MR-curves taken at different temperatures scale according to
Kohler's rule, implying a nonmagnetic scattering process. Vanadium deficient
samples show a negative contribution to the MR which increases with increasing
$y$, indicating the formation of magnetic moments. In the SDW phase Kohler's
rule is not obeyed and magnetic scattering dominates the MR.

\begin{figure}[bt]
 \begin{center}
  \epsfig{file=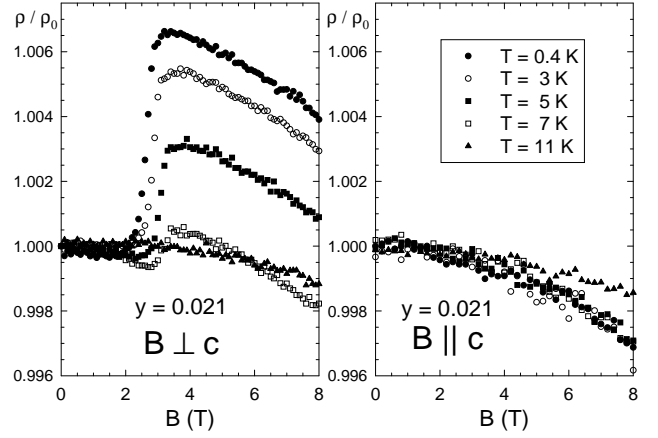, width=\columnwidth}
 \end{center}
 \caption{Magnetoresistance of a sample in the SDW~1 phase at $B=0$.}
 \label{MR_type1}
\end{figure}

In the SDW~1 phase, i.e.\ for high doping levels, the resistivity for ${\bf
B}\parallel{\bf c}$ decreases monotonously with increasing $B$
(Fig.~\ref{MR_type1}, right). For a field applied perpendicular to the
$c$-axis, however, a dramatically different behavior is observed as shown in
the left part of Fig.~\ref{MR_type1}. At a critical field $B$ the resistivity
suddenly jumps to a higher value. For clarity only curves with increasing field
are shown. On lowering the field below the critical value a hysteresis of about
$0.1$~T is observed. Since a corresponding anomaly was observed in the
magnetization (Fig.~\ref{magnetization}) we also label the critical field
inferred from the MR anomaly as $B_{c,\perp}$. With increasing temperature
$B_{c,\perp}$ increases slightly while the size of the jump decreases until it
vanishes completely for $T>T_{SDW}$. We attribute this metamagnetic behavior to
the transition from SDW~1 to SDW~2, as was already inferred from magnetization
measurements. The critical field $B_{c,\perp}$ increases with increasing doping
at a slope of $dB_{c,\perp}/dy \approx 4 \times 10^2$~T.

\begin{figure}[bt]
 \begin{center}
  \epsfig{file=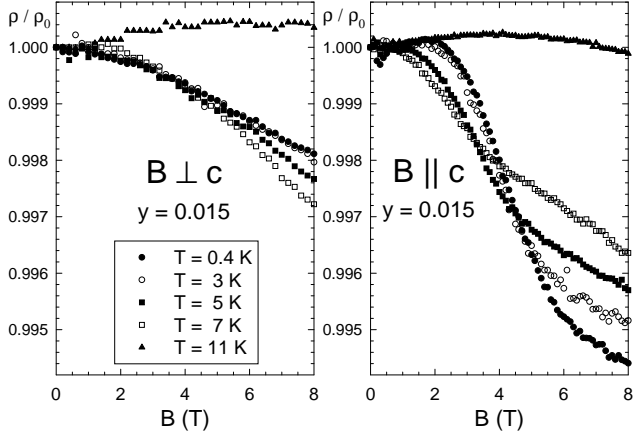, width=\columnwidth}
 \end{center}
 \caption{Magnetoresistance of a sample in the SDW~2 phase at $B=0$.}
 \label{MR_type2}
\end{figure}

For a sample with composition close to the AFI phase boundary ($y=0.015$) the
MR behaves completely differently (Fig.~\ref{MR_type2}). Here a sudden jump to
lower resistivity values occurs, with the magnetic field directed along the
$c$-axis. The corresponding critical field, taken as the inflection point of
this anomaly, decreases with increasing temperature. This critical field has
$B_{c,\parallel}$ as its counterpart. For ${\bf B}\perp{\bf c}$ no anomaly
exists. This behavior, which we attribute to the SDW~2 phase, persists in this
sample for all temperatures up to $T_{SDW}=11$~K.

\begin{figure}[bt]
 \begin{center}
  \epsfig{file=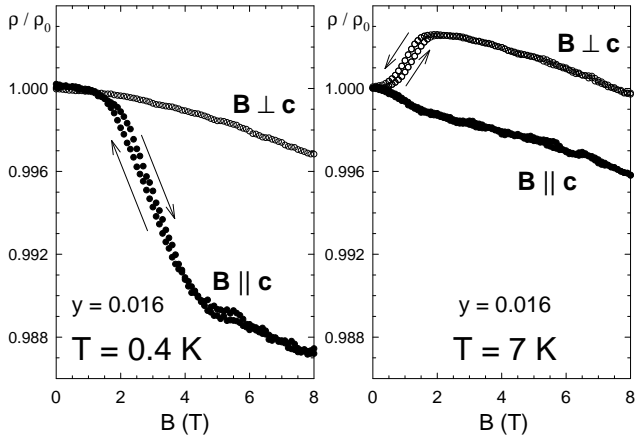, width=\columnwidth}
 \end{center}
 \caption{MR of the $y=0.016$ sample with transition from SDW~2 to SDW~1 at
  low  temperatures (left) and transition from SDW~1 to SDW~2 at higher
  temperatures (right).}
 \label{MR_T12}
\end{figure}

For slightly higher doping ($y \approx 0.016$) a transition from SDW~1 to SDW~2
phase as a function of temperature is observed (Fig.~\ref{MR_T12}) in
accordance with our findings from susceptibility measurements
(Fig.~\ref{chi_316}).
The transition between the two different behaviors of the MR, characteristic of
the two different SDW phases, occurs around $T\approx 6$~K. The characteristic
field $B_{c,\parallel}$ at lowest temperature is smaller for the $y=0.016$
sample than for the $y=0.015$ sample. Both magnetic field induced transitions,
(SDW~1 to SDW~2 and SDW~2 to SDW~1) exhibit a hysteresis shown in
Fig.~\ref{MR_T12} which was omitted for clarity in Fig.~\ref{MR_type1} and
Fig.~\ref{MR_type2}.

\begin{figure}[bt]
 \begin{center}
  \epsfig{file=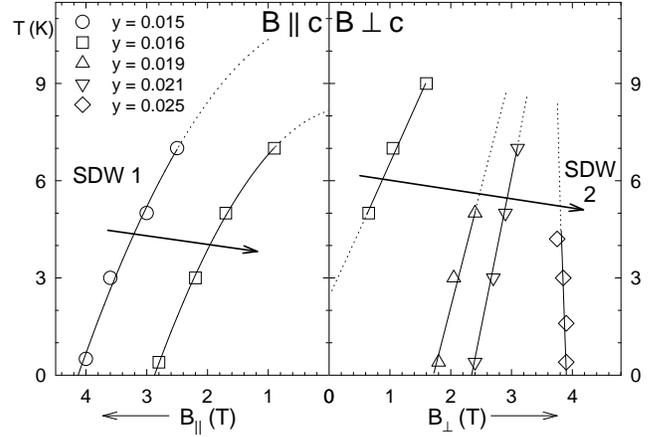, width=\columnwidth}
 \end{center}
 \caption{$T$--$B$ diagram of the phase boundary between SDW~1 and SDW~2
  for ${\bf B}\parallel{\bf c}$ (left) and ${\bf B}\perp{\bf c}$ (right). The
  arrows indicate the direction in which the phase boundary shifts with
  increasing doping.}
 \label{T_B_pd}
\end{figure}

The direction of the magnetic field has to be considered for the phase boundary
between SDW~1 and SDW~2 in a $T$~--~$B$ phase diagram, i.e.\ separate diagrams
for the two field directions have to be plotted. This is illustrated in
Fig.~\ref{T_B_pd} for various samples of different stoichiometry.
The symbols correspond to $B_{c,\perp}$ (right) and $B_{c,\parallel}$ (left),
respectively, as determined from the inflection point of the MR curves. Note
that the $B_{c,\parallel}$-axis is reversed to account for the opposite effect
of $B_\perp$ and $B_\parallel$ on the stability of the two phases. For the
$y=0.015$ sample the SDW~1~--~SDW~2 boundary lies completely in the ${\bf
B}\parallel{\bf c}$ part of the phase diagram, i.e.\ the SDW~2 phase is stable
in zero field as well as for all values of $B_\perp$. The SDW~1 phase can be
induced only for large $B_\parallel$. On increasing $y$ this phase boundary
shifts to the ${\bf B}\perp{\bf c}$ side of the phase diagram as indicated by
the arrows until above $y\approx 0.017$ the SDW~1 phase is stable down to the
lowest temperatures in zero field. Then SDW~2 can only be induced by a
sufficiently high $B_\perp$ which increases with increasing doping. At the same
time the phase boundary becomes steeper until for $y=0.025$ $dB_{c,\perp}/dT$
even changes sign. Unfortunately a sample with higher doping was not available.
The $y=0.04$ sample investigated by Bao {\it et al.} \cite{Bao01}, however,
shows a negative $dB_{c,\perp}/dT$ confirming the general trend observed in our
study. For the $y=0.016$ sample, which crosses the phase boundary at zero field
(square symbols in Fig.~\ref{T_B_pd}), the phase boundaries for ${\bf
B}\parallel{\bf c}$ and ${\bf B}\perp{\bf c}$ do not match at $B=0$. This is
due to the large hysteresis on cycling the temperature as already shown in the
susceptibility (Fig.~\ref{chi_316}). In the temperature range 4~K$<T<$8~K both
SDW phases apparently coexist in this sample and finite fields $B_{c,\perp}$ as
well as $B_{c,\parallel}$ are needed to stabilize pure SDW~2 or SDW~1 phase,
respectively.

\subsection{Pressure dependence}

To clarify whether the transition between SDW~1 and SDW~2 depends only on the
hole concentration, the effect of external pressure, which does not change the
band filling, should be examined. In general, pressure appears to have the same
effect as doping on the V$_{2-y}$O$_3$ phase diagram. Starting from the
stoichiometric compound at ambient pressure, the reduction of $T_{MI}$ by
doping and pressure scales with a ratio of $\Delta p/\Delta y \approx 1.4\times
10^2$~GPa. However, in contrast to the doping induced metallic phase, the
pressure induced metallic phase of the stoichiometric sample does not show a
transition into a SDW phase down to $T=0.3$~K (see Fig.~\ref{3D}). If pressure
is applied to samples exhibiting a SDW ground state, $T_{SDW}$ is reduced until
the PM phase is stabilized down to the lowest temperatures. The sensitivity of
the SDW ordering temperature to pressure $(dT_{SDW}/dp)$ decreases with
increasing doping although $T_{SDW}$ decreases. Thus, except for the absence of
a SDW phase in the pressure induced metallic state of the stoichiometric
sample, the effects of pressure and doping are similar.

\begin{figure}[bt]
 \begin{center}
  \epsfig{file=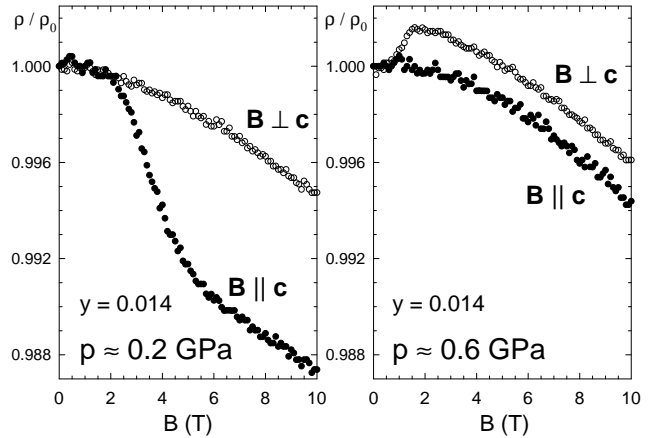, width=\columnwidth}
 \end{center}
 \caption{MR curves at $T=0.4$~K for two different pressures showing the
  transition from SDW~2 behavior (left) to SDW~1 behavior (right).}
 \label{MR-p}
\end{figure}

We performed pressure dependent magnetotransport measurements on seven samples
with different levels of vanadium deficiency. In the following we only want to
discuss information from these measurements relevant to the phase diagram of
V$_{2-y}$O$_3$. For the stoichiometric sample (already discussed above) and for
$y=0.003$, pressure induces a transition from the AFI phase to a paramagnetic
metallic ground state. The samples with $y=0.09$ and $y=0.014$ show a more
complex behavior. This is illustrated in Fig.~\ref{MR-p} for the $y=0.014$
sample.
At ambient pressure this sample assumes an AFI ground state
($T_{MI}=50$~K). At a pressure of 0.1~GPa the sample remains metallic down to
lowest temperatures. At this and slightly higher pressures we observe a
behavior characteristic for the SDW~2 phase (Fig.~\ref{MR-p}, left), as it was
found, e.g., for the $y=0.015$ sample (Fig.~\ref{MR_type2}) at $p=0$. At a
pressure of $p=0.4$~GPa both SDW~1 and SDW~2 behavior can be observed,
depending on the temperature. This behavior is reminiscent of the $y=0.016$
sample, which shows a transition from SDW~1 to SDW~2 as a function of
temperature. Increasing the pressure further to 0.6~GPa results in MR features
characteristic of the SDW~1 phase (Fig.~\ref{MR-p}, right). The critical field
$B_{c,\perp}$, that induces the SDW~2 phase, increases with increasing
pressure. Finally, at $p=2$~GPa no magnetic ordering is observed down to the
lowest temperatures.

Hence, the ground state of this sample undergoes three phase transitions as a
function of pressure, AFI to SDW~2 phase, SDW~2 to SDW~1 phase and, finally,
SDW~1 phase to a paramagnetic metal. For a fixed pressure of 0.4~GPa three
phases are observed (SDW~2~--~SDW~1~--~PM) as a function of temperature.

\begin{figure}[bt]
 \begin{center}
  \epsfig{file=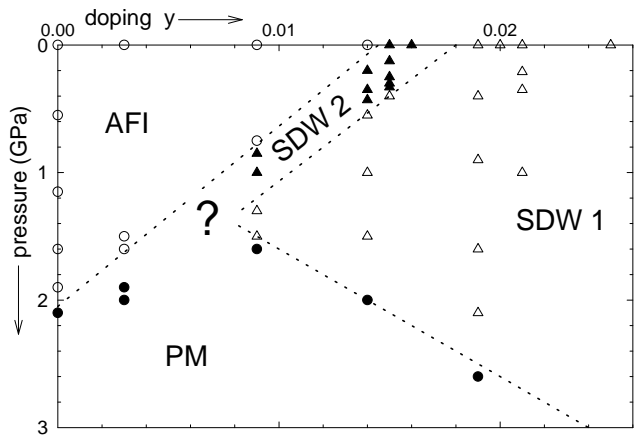, width=\columnwidth}
 \end{center}
 \caption{Schematic phase diagram in the $p$~--~$y$ plane. The symbols denote
  the $y$ of the various samples and the respective pressure applied. Open
  round symbols indicate an AFI ground state, closed round symbols the
  paramagnetic metallic phase, closed triangles the SDW~2 and open triangles
  the SDW~1 phase. All phase boundaries are approximated by straight lines.}
 \label{p-y-pd}
\end{figure}

For the sample with $y = 0.019$, which shows the SDW~1 phase ground state,
pressure induces the paramagnetic metallic (PM) phase. The pressure dependent
measurements are summarized in the $p$~--~$y$ ground state phase diagram
($T=0.4$~K) in Fig.~\ref{p-y-pd}.
The newly discovered SDW~2 phase exists only in a small range of doping and
pressure close to the AFI phase. Within experimental resolution the
SDW~1~--~SDW~2 and SDW~2~--~AFI phase boundaries appear to be parallel. The
situation at the SDW~2~--~PM boundary remains unclear (question mark in
Fig.~\ref{p-y-pd}), since no samples with suitable doping were available. The
$p$~--~$y$ diagram corresponds to the base of Fig.~\ref{3D} while the rear wall
($T$~--~$y$ diagram) was already shown in Fig.~\ref{T-y-pd}.

\section{Discussion}

From our susceptibility and magnetotransport data it can be concluded that two
SDW phases exist. Since the different phases of V$_{2-y}$O$_3$ can be observed
in a sample with given stoichiometry as a function of pressure we do not
believe that disorder introduced by the vanadium deficiency is responsible for
the properties of this compound. While the SDW phase at large $y$ (SDW~1) has
been extensively studied by neutron scattering \cite{Bao93}, the exact nature
of SDW~2 remains unclear. However, in the study of Bao {\it et al.}
\cite{Bao93} an additional incommensurate peak corresponding to a slightly
larger {\bf q} vector was observed at temperatures below 6~K for one sample
with a composition close to the AFI~--~SDW phase boundary. This finding has so
far been denoted "unclear". Taking our new results into account it appears
evident that this peak signals the presence of the SDW~2 phase. Since hard and
easy axes seem to be interchanged between SDW~1 and SDW~2 the moments are
expected to be oriented along the $c$-axis or at least have an appreciable out
of plane component in SDW~2. Unpublished neutron studies of this second phase
confirm this assumption \cite{Bao-p}. Most likely, the SDW~2 phase is a
longitudinal SDW with propagation vector still along the $c$-axis, but with the
$c$-component of the moments being modulated. Due to this antiferromagnetic
coupling of the $c$-component it is hard to induce a magnetization by a field
along the $c$-axis in SDW~2, while in SDW~1 the antiferromagnetic exchange
couples the in plane component of the moments and so a magnetization along the
$c$-axis is more easily obtained. For the same reason, a field applied
perpendicular to the $c$-axis aligns the moments more easily in SDW~2 than in
SDW~1. Thus in a sufficiently large field along the $c$-axis the free energy is
lower in SDW~1 compared to SDW~2 and vice versa for ${\bf B}\perp{\bf c}$. The
assumption of a longitudinal polarization in SDW~2 therefore gives a consistent
interpretation of the observed anisotropic response to a magnetic field for
both phases.

The MR data show the resistivity is somewhat higher in the SDW~2 than in the
SDW~1 phase indicating that an increased portion of the Fermi surface is
removed by the formation of SDW~2. This results in a reduced density of free
charge carriers and a lower entropy of SDW~2 compared to SDW~1. Thermodynamic
considerations therefore predict the SDW~1 phase to become more stable compared
to SDW~2 with increasing temperature consistent with the experimentally
observed transition in the $y=0.016$ sample and the corresponding phase diagram
(Fig.~\ref{T-y-pd}).

Discussing the stability of the different phases with respect to the parameters
doping and pressure it is interesting to note, that both external pressure and
doping can induce the transition from SDW~2 into SDW~1. Since the SDW formation
originates from Fermi surface nesting \cite{Wolenski98}, it follows, that both
doping and pressure induces similar changes in the Fermi surface. Band
structure calculations \cite{Mattheiss94,Ezhov99,LDA_c/a} predict the bands
near $E_F$ to be of a$_{1g}$ and e$_g^\pi$ character. Most likely a band with
predominant a$_{1g}$ character, which has large dispersion along the
$c$-direction, provides the best nesting conditions and therefore should be
responsible for the formation of the SDW.

Hole doping $y$ reduces the band filling. A vanadium deficiency of 1\%
($y=0.02$) changes the filling of the t$_{2g}$ bands by 2.5\%. In addition, a
volume contraction of $\Delta V/V \approx 0.3$\% at $y=0.02$ occurs at room
temperature, while the $c/a$ ratio remains almost constant \cite{Ueda80}. On
the other hand, external pressure does not change the overall band filling, but
also causes a volume contraction (of 1\% at 2~GPa) accompanied by an 0.35\%
increase of the $c/a$ ratio \cite{Sato79,Finger80}. Although there is only
scanty structural data at low temperatures, it can be assumed that similar
changes of volume and corresponding changes of $a$- and $c$-axis parameters
occur as a function of doping and pressure in the SDW regime. It should also be
noted that there is a substantial increase of the $c/a$ ratio with decreasing
temperature of about 0.4\% between RT and $T = 77$~K \cite{Ueda80}.

In a simplistic band picture, the relative weight of a$_{1g}$ to e$_g^\pi$
character can be changed by depopulating one of the bands by hole doping or
shifting it to higher energies by pressure. In fact, a sensitivity of band
positions in the vicinity of the Fermi energy to pressure and the according
structural changes might be expected from changes of the V$3d$--O$2p$
hybridization. Such changes of hybridization are observed at the PM--AFI
transition for stoichiometric samples \cite{Mueller97}. In addition, dramatic
changes of the relative position of the different bands at the Fermi level have
been found in recent band structure calculations \cite{LDA_c/a} varying the
$c/a$ ratio within the pseudo hexagonal PM phase, resulting in substantial
variations of the corresponding Fermi surfaces. Further support for a possible
shift of relative band positions of bands with a$_{1g}$ and e$_g^\pi$ character
in V$_2$O$_3$ is provided by a similar, much more pronounced effect in the
isostructural compound Ti$_2$O$_3$ \cite{Mattheiss96}. We therefore believe,
that relative shifts of bands of a$_{1g}$ and e$_g^\pi$ character crossing the
Fermi energy are responsible for the observed transition between two different
SDW states. In this context we note, that such shifts could also effect the
other phases in the complex V$_2$O$_3$ phase diagram, which are, of course,
rendered even more complex by the presence of electronic correlations.

\section{Conclusion}

We have shown that, depending on vanadium deficiency $y$, two different SDW
ground states exist in V$_{2-y}$O$_3$. The transition between the two SDW
states as a function of doping and pressure was attributed to changes of the
Fermi surface resulting from band filling and structurally induced band shifts,
respectively. From magnetic susceptibility and magnetoresistance measurements,
the latter performed as a function of pressure, tentative phase diagrams were
deduced with parameters temperature, magnetic field, pressure, and doping.

\acknowledgments

This work was supported by the Deutsche Forschungsgemeinschaft under contract
number HO~955/2-1 and SFB 484. We are grateful to P. Riseborough for valuable
discussions and to V. Eyert for providing the LDA band structure calculations
before publication.

\end{document}